\documentclass[10pt,twocolumn, conference]{IEEEtran}

\author{\authorblockN{Qingyun Wang \hspace{0.5cm}
Mustafa Cenk Gursoy}
\authorblockA{Department of Electrical Engineering\\
University of Nebraska-Lincoln, Lincoln, NE 68588\\ Email:
qwang4@bigred.unl.edu, gursoy@unl.edu}}

\usepackage{amsmath}
\usepackage{amssymb}
\usepackage[dvips]{graphicx}
\usepackage{setspace}
\usepackage{epsfig}

\newcommand{\h}{\mathbf{h}}

\newcommand{\R}{{\mathbf{R}}}

\newcommand{\tth}{{^\text{th}}}

\newtheorem{lemma1}{Lemma}

\begin{document}

\title{Performance Analysis for Multichannel Reception of OOFSK Signaling}

\date{}

\maketitle

\begin{abstract}\footnote{This work was supported in part by the NSF CAREER
Grant CCF-0546384.} In this paper, the error performance of on-off
frequency shift keying (OOFSK) modulation over fading channels is
analyzed when the receiver is equipped with multiple antennas. The
analysis is conducted in two cases: the coherent scenario where the
fading is perfectly known at the receiver, and the noncoherent
scenario where neither the receiver nor the transmitter knows the
fading coefficients. For both cases, the maximum a posteriori
probability (MAP) detection rule is derived and analytical
probability of error expressions are obtained. The effect of fading
correlation among the receiver antennas is also studied. Simulation
results indicate that for sufficiently low duty cycle values, lower
probability of error values with respect to FSK signaling are
achieved. Equivalently, when compared to FSK modulation, OOFSK with
low duty cycle requires less energy to achieve the same probability
of error, which renders this modulation a more energy efficient
transmission technique.
\end{abstract}

\section{Introduction}

Frequency-shift keying (FSK) is a modulation format that is
well-known and well-studied in the communications literature
\cite{proakis}. FSK is an attractive transmission scheme due to its
high energy efficiency and suitability for noncoherent
communications. In unknown channel conditions, energy detection can
be employed to detect the FSK signals. Indeed, the analysis of FSK
modulation dates back to 1960s (see e.g., \cite{Pierce}, and
\cite{lindsey}). Recently, it has been shown in \cite{Verdu} that
unless the channel conditions are perfectly known at the receiver,
signals that have very high peak-to-average power ratio is required
to achieve the capacity in the low SNR regime. This has initiated
work on peaky signaling. Luo and M\'edard \cite{LuoMedard} have
shown that FSK with small duty cycle can achieve rates of the order
of capacity in ultrawideband systems with limits on bandwidth and
peak power. In \cite{desmond}, the authors have studied the error
performance of peaky FSK signaling over multipath fading channels by
obtaining upper and lower bounds on the error probability. In
\cite{gursoyoofsk}, on-off frequency-shift keying (OOFSK) is defined
as FSK overlaid on on-off keying, and its capacity and energy
efficiency is analyzed. Note that OOFSK can be seen as joint pulse
position modulation (PPM) and FSK. In this signaling, peakedness is
introduced in both time and frequency. The error performance of
OOFSK signaling when the transmitter and receiver are each equipped
with a single antenna is recently studied in \cite{qingyun}.

One of the important techniques to improve the performance in
wireless communications is to use multiple antennas to achieve
diversity gain. Considerable amount of work has been done on
multiple reception channels. In \cite{lindsey}, it is shown for
binary and $M$-ary signaling over Rician fading channels that
increasing the number of reception channels can improve the error
performance significantly. By finding the probability distribution
function of the instantaneous SNR in flat fading multi-reception
channels and substituting it into the probability of error
expressions of PAM, PSK and QAM over AWGN channel, the authors in
\cite{hao} obtained expressions for the average probability of error
of multi-reception fading channels. In \cite{veer}, the probability
of error of BPSK over Rician fading multi-reception channels is
given and extensions to other modulation techniques are discussed.
In \cite{yunfei}, average symbol error rate of selection diversity
of $M$-ary FSK modulated signal transmitted over fading channels is
studied.

In this paper, the error performance of OOFSK over multiple
reception Rician fading channels is studied. In Section
\ref{sec:model}, the system model is presented. In Section
\ref{sec:coherent}, the error performance in coherent fading
channels is studied. In Section \ref{sec:noncoherent}, we
investigate the error performance in noncoherent Rician fading
channels.

\section{System Model}\label{sec:model}

We assume that OOFSK modulation is employed at the transmitter to
send the information. In OOFSK modulation, the transmitted signal
during the symbol interval $0 \le t \le T_s$ can be expressed as
\begin{align}
s_m(t)=\left\{\begin{array}{ll}\sqrt{\frac{P}{v}}e^{j(w_mt+\theta_m)}
& m=1,2,3,\ldots M\\0  & m=0
\end{array}\right.
\end{align}
where $w_m$ and $\theta_m$ are the frequency in radians per second
and phase, respectively, of the signal $s_m(t)$ when $m \neq 0$.
Note that we have $M$ FSK signals and a zero signal denoted by
$s_0(t)$. The frequencies of the FSK signals are chosen so that the
signals are orthogonal. It is assumed that an FSK signal $s_m(t)$,
$m \neq 0$, is transmitted with a probability of $\frac{v}{M}$ while
$s_0(t)$ is transmitted with a probability of $1-v$ where $v$ is the
duty cycle of transmission. With these definitions, it is easily
seen that $P$ and $\frac{P}{v}$ are the average and peak powers,
respectively, of the modulation technique.
\par The receiver is equipped with $L$ antennas that enable
the multiple reception of the transmitted signal. If, without loss
of generality, we assume that $s_k(t)$ is the transmitted signal,
the received signal at the $l^{\text{th}}$ antenna is
\begin{equation} \label{eq:receivedsignal}
r_l(t)=h_ls_k(t)+n_l(t)\quad l=1,2,\ldots,L
\end{equation}
where $h_l$ is the fading coefficient at the  $l^{\text{th}}$
reception channel and $n_l(t)$ is a white Gaussian noise with
single-sided spectral density of $N_0$. It is assumed that the
additive Gaussian noise components at different antennas are
independent. Furthermore, the received signal model
(\ref{eq:receivedsignal}) presumes that the fading is frequency-flat
and slow enough so that the fading coefficient stays constant over
one symbol duration.

Following each antenna, there is a bank of $M$ correlators, each
correlating the received signal with one of the orthogonal
frequencies.
The output of the $m^{\text{th}}$ correlator
employed after the $l^{\text{th}}$ antenna is given by
\begin{align}\label{eq:output}
Y_{l,m}&=\frac{1}{\sqrt{N_0T_s}}\int_{0}^{T_s}r_l(t)e^{-jw_mt}dt\nonumber\\
&=\left\{\begin{array}{ll} \sqrt{\frac{PT_s}{vN_0}}h_l
e^{j\theta_m}+n_{l,m} & m=k  \\
n_{l,m} &m\neq k \end{array}\right.\nonumber \\
&=\left\{\begin{array}{ll}A h_l
e^{j\theta_m}+n_{l,m} &m=k\\
n_{l,m} &m\neq k, \end{array}\right.
\end{align}
where $n_{l,m}$ is a circularly symmetric complex Gaussian random
variable with zero-mean and a variance of $1$ and for notational
convenience, we have defined $A=\sqrt{\frac{PT_s}{vN_0}}$. Since the
frequencies are orthogonal and the additive Gaussian noise is
independent at each antenna, $\{n_{l,m}\}$ for $l \in
\{1,\ldots,L\}$ and $m \in \{1,\ldots,M\}$ forms an independent and
identically distributed (i.i.d.) sequence. Note also that
$R_{l,m}=|Y_{l,m}|^2$ gives the energy present in the
$m^{\text{th}}$ frequency at the $l^{th}$ antenna.

\section{OOFSK Over Coherent Fading Channels} \label{sec:coherent}

\subsection{Detection Rule}
In this section, we assume that transmission takes place over
coherent fading channels and hence $h_l$ for all $l$ is known to the
receiver while the transmitter does not have such knowledge.
Conditioned on $h_l$ and the transmitted signal $s_k(t)$, $Y_{l,m}$
is a proper complex Gaussian random variable with mean value and
variance given by
\begin{align}
&E\{Y_{l,m}|h_l,s_k\}=\left\{\begin{array}{ll}Ah_le^{j\theta_k}  & m=k\\
0 & m\neq k
\end{array}\right.
\\
&var\{Y_{l,m}|h_l,s_k\}=1.
\end{align}
Therefore, $R_{l,m} = |Y_{l,m}|^2$ is chi-square distributed with
the following conditional probability density function (pdf):
\begin{small}
\begin{align}
f_{R_{l,m}|\,|h_l|,s_k}(R_{l,m})\!\!=\!\!\left\{\begin{array}{ll}\!\!\!\!e^{-(R_{l,m}+A^2|h_l|^2)}I_0\left(2A|h_l|\sqrt{R_{l,m}}\right)&m=k\\
\!\!\!\!e^{-R_{l,m}}&m\neq k\end{array}\right. \nonumber
\end{align}
\end{small}
It is assumed that the receiver, using equal gain combining (EGC),
combines the energies of the $m^{\text{th}}$ frequency components
at each antenna, i.e., computes the total energy
\begin{gather}
R_m=\sum_{l=1}^LR_{l,m}.
\end{gather}
Since the noise components are independent, $R_m$ is a sum of
independent chi-square random variables, and is itself also
chi-square distributed with $2L$ degrees of freedom. The conditional
pdf is given by \begin{small}
\begin{align}
f_{R_m| \,
\h,s_k}(R_m)=\left\{\begin{array}{ll}\left(\frac{R_m}{\xi}\right)^{\frac{L-1}{2}}e^{-(R_m+\xi)}I_{L-1}(2\sqrt{R_m\xi})
&m=k\nonumber \vspace{.1cm}\\
\frac{R_m^{L-1}}{\Gamma(L)}e^{-R_m}&m\neq k
\end{array}\right.
\end{align}
\end{small}
where $\xi=\sum_{l=1}^LA^2|h_l|^2$, $\h = [h_1,\ldots,h_L]$,
$I_{L-1}(\cdot)$ is the $(L-1)^{\text{th}}$ order modified Bessel
function of the first kind, and $\Gamma(\cdot)$ is the gamma
function.

The receiver employs maximum a posteriori probability (MAP)
criterion to detect the transmitted signals.  Let
$$\R=[R_1,R_2,\ldots,R_M]$$ be the vector of energy values corresponding to
each
frequency. Since the noise components $n_{l,m}$ are independent for
different $m \in \{1,\ldots,M\}$, components of $\R$ are mutually
independent. Hence, the conditional pdf of $\R$ is {\small{
\begin{align}\label{pdf_sm}
&f_{\R|\h, s_k}(\R)= \nonumber \\
&\left\{\begin{array}{ll}\!\!\!\!\left(\frac{R_k}{\xi}\right)^{\frac{L-1}{2}}e^{-(R_k+\xi)}I_{L-1}(2\sqrt{R_k\xi})\prod_{n=1\
n\neq k
}^M \frac{R_n^{L-1}e^{-R_n}}{\Gamma(L)} &  k\neq 0\\
\!\!\!\frac{1}{\left[\Gamma(L)\right]^{M}}\prod_{n=1}^MR_n^{L-1}e^{-R_n}
& k=0\end{array}\right.
\end{align}}}
Then, the MAP rule that detects $s_k$ for $k \neq 0$ is
\begin{align}\label{compare1}
\left\{\begin{array}{ll}f_{\R|\h, s_k}>f_{\R|\h, s_m} & \forall m\neq 0, k
\\
f_{\R|\h, s_k}>\frac{M(1-v)}{v}f_{\R|\h, s_0}\end{array}\right.
\end{align}
where we have used the fact that the prior probabilities of the
transmitted signals are $p(s_m) = \frac{v}{M}$ for $m \neq 0$, and
$p(s_0) = (1-v)$. Substituting (\ref{pdf_sm}) into to
(\ref{compare1}), the decision rule is simplified to:
\begin{align}\label{com1}
\left\{\begin{array}{ll}g_1(R_k)>g_1(R_m) &\forall m \neq k\\
g_1(R_k)>\frac{M(1-v)e^\xi
\xi^{\frac{L-1}{2}}}{v(L-1)!}\end{array}\right.
\end{align}
where
\begin{gather}\label{eq:functiong}
g_1(R_k) = R_k^{-\frac{L-1}{2}}I_{L-1}(2\sqrt{R_k\xi}),\quad \xi>0.
\end{gather}
The following
Lemma enables us to further simplify the detection rule.
\begin{lemma1}
The function
\begin{gather}
g_1(x)=x^{-\frac{L-1}{2}}I_{L-1}(2\sqrt{x\xi}) \quad \text{for } x >
0, \xi
> 0
\end{gather}
is a monotonically increasing function of $x$.
\end{lemma1}

\emph{Proof}: The derivative of the $n\tth$ order modified Bessel
function is
\begin{gather}
\frac{dI_n(x)}{dx} = I_{n+1}(x) + \frac{n}{x}I_n(x).
\end{gather}
Hence,
\begin{align}
\frac{d I_{L-1}(2\sqrt{x\xi})}{dx} &=
\sqrt{\frac{\xi}{x}}I_L(2\sqrt{x\xi}) +
\frac{L-1}{2x}I_{L-1}(2\sqrt{x\xi})\nonumber
\\
&> \frac{L-1}{2x}I_{L-1}(2\sqrt{x\xi})
\end{align}
 where we use the fact that $\sqrt{\frac{\xi}{x}}I_L(2\sqrt{x\xi})>0$ for $x>0$. Then, the derivative of $g_1(\cdot)$ satisfies
\begin{align}
&\frac{dg_1(x)}{dx}=-\frac{L-1}{2}x^{-\frac{L+1}{2}}I_{L\!-\!1}(2\sqrt{x\xi})+x^{-\frac{L-1}{2}}\frac{dI_{L-1}(2\sqrt{x\xi})}{dx}\nonumber\\
&>
-\frac{L-1}{2}x^{-\frac{L+1}{2}}I_{L\!-\!1}(2\sqrt{x\xi})+x^{-\frac{L-1}{2}}
\frac{L-1}{2x}I_{L-1}(2\sqrt{x\xi})\nonumber\\
&=0, \nonumber
\end{align}
proving that $g_1(x)$ is a monotonically increasing function of
$x>0$. \hfill $\blacksquare$

By the above result, the detection rule (\ref{com1}) further
simplifies to
\begin{align}\label{finalcompare1}
\left\{\begin{array}{ll}R_k>R_m & \forall m \neq k\\
R_k>g_1^{-1}(T)\end{array}\right.
\end{align}
where $T=\frac{M(1-v)e^\xi \xi^{\frac{L-1}{2}}}{v(L-1)!}$. Since
$g_1(\cdot)$ is a monotonically increasing function, the inverse
$g_1^{-1}(\cdot)$ is well-defined. Note that (\ref{finalcompare1})
is the rule that detects the signal $s_k(t)$ for $k \neq 0$. The
zero signal $s_0(t)$ is detected if
\begin{align}
R_k < g_1^{-1}(T) \quad \forall k.
\end{align}

\subsection{Probability of Error}
In this section, we analyze the error probability of OOFSK
modulation when MAP detection is used at the receiver. Suppose
without loss of generality that $s_1(t)$ is the transmitted signal.
Let $\tau=g_1^{-1}(T)$. Then the correct detection probability is
\begin{align}
P_{c,1}&=P(R_1>R_2,R_1>R_3,\ldots,R_1>R_M,R_1>\tau|s_1)\nonumber\\
&=\int_\tau^\infty\left(\int_0^{x}f_{R_2|\h,
s_1}(t)\,dt\right)^{M-1}f_{R_1|\h, s_1}(x)\,dx\nonumber\\
&=\int_\tau^\infty\left(\int_0^{x}\frac{t^{L-1}}{\Gamma(L)}e^{-t}dt\right)^{M-1}f_{R_1|\h,
s_1}(x)\,dx\nonumber.
\end{align}
From \cite{proakis}, we have
 \begin{equation}
\int_0^x\frac{1}{\Gamma(L)}t^{L-1}e^{-t}dt=1-e^{-x}\sum_{l=0}^{L-1}\frac{x^l}{l!}
 \end{equation}
Therefore, the correct detection probability can now be expressed as
\begin{align}
P_{c,1}&=\int_\tau^\infty\left[1-e^{-x}\sum_{l=0}^{L-1}\frac{x^l}{l!}\right]^{M-1}f_{R_1|
\h, s_1}(x)dx\nonumber
\end{align}
Using the binomial theorem, $P_{c,1}$ becomes
\begin{align}
P_{c,1}&=\!\!\int_\tau^\infty\!
\sum_{n=0}^{M\!-\!1}(-1)^n\!\!\left(\!\!\!\begin{array}{cc}M\!-\!1\\n\end{array}\!
\!\!\right)\!\!\!\left[\sum_{l=0}^{L-1}\!\!\frac{x^l}{l!}e^{-x}\right]^n\!\!\!f_{R_1|
\h, s_1}(x)dx\nonumber
\end{align}
Using the multinomial theorem, we have the following expansion
\begin{align}\label{multinomial}
\left[\sum_{l=0}^{L-1}\frac{x^l}{l!}e^{-x}\right]^n=e^{-nx}\sum_{i=0}^{n(L-1)}c_{in}x^i
\end{align}
where $c_{in}$ is the coefficient of $x^i$ in the expansion.
$c_{in}$ can be evaluated from the recursive equation \cite{marvin}
\begin{equation}
c_{in}=\sum_{q=i-L+1}^i\frac{c_{q(n-1)}}{(i-q)!}I_{[0,(n-1)(L-1)]}(q)
\end{equation}
where
\begin{align}
I_{[a,b]}(q)=\left\{\begin{array}{ll} 1, \quad a\leq q\leq b\\
0, \quad \text{otherwise} \end{array} \right..
\end{align}
Using the multinomial expansion, $P_{c,1}$ becomes
\begin{align}
P_{c,1}&=\!\!\sum_{n=0}^{M-1}(-1)^n\left(\!\!\!\begin{array}{cc}M-1\\n\end{array}\!\!\!\right)\sum_{i=0}^{n(L-1)}c_{in}\int_{\tau}^\infty
x^i\left(\frac{x}{\xi}\right)^{\frac{L-1}{2}}\nonumber\\
&\quad\times e^{-[(n+1)x+\xi]}I_{L-1}(2\sqrt{x\xi})dx
\end{align}
Let $\xi=a^2$ and $x=t^2$, then, $P_{c,1}$ can be written as
\begin{align}
P_{c,1}\!\!&=\!\!\sum_{n=0}^{M\!-\!1}(-1)^n\left(\!\!\!\!\begin{array}{cc}M-1\\n\end{array}\!\!\!\!\right)\sum_{i=0}^{n(L\!-\!1)}\!\!c_{in}\nonumber\\
&\quad\times\int_{\sqrt{\tau}}^\infty
t^{2i}e^{-nt^2}\left(\frac{t}{a}\right)^{L-1}e^{-(t^2+a^2)}I_{L-1}(2at)2tdt\nonumber\\
&=\sum_{n=0}^{M-1}(-1)^n\left(\begin{array}{cc}M-1\\n\end{array}\right)\sum_{i=0}^{n(L-1)}2c_{in}e^{a^2}a^{-(L-1)}\nonumber\\
&\times\left[\frac{a^{L-1}\Gamma(i+L)}{2(n+1)^{i+L}\Gamma(L)}\, e^{\frac{a^2}{n+1}}F\left(-i,L;\frac{-a^2}{n+1}\right) \right.\nonumber\\
&\,\,\,\,\,\,\,\,\,\,\left.-\int_0^{\sqrt{\tau}}
\!\!\!\!t^{2i+L}e^{-(n+1)t^2}I_{L-1}(2at)dt\right]
\end{align}
where $F(a,c;x)$ is the confluent hypergeometric function
\cite[Chap 10]{andrews}.
The probability of correct detection when signal $s_0(t)$ is
transmitted is:
\begin{align}
P_{c,0}&=P\left(R_1<\tau,\ldots,R_M<\tau|\h,s_0\right)\nonumber\\
&= \left( 1 - e^{-\tau} \sum_{l=0}^{L-1} \frac{\tau^l}{l!}\right)^M.
\end{align}
Hence, the probability of error as a function of the instantaneous
signal-to-noise ratio is
\begin{align} \label{eq:proboferror}
P_e=1-(vP_{c,1}+(1-v)P_{c,0}).
\end{align}
Since the channel is assumed to be known, error probability in
(\ref{eq:proboferror}) is a function of the fading coefficients
through $\chi=\sum_{l=1}^{L}|h_l|^2$. Hence, the average probability
of error is obtained by computing
\begin{align}
\bar{P}_e&=\int_0^\infty P_e f_{\chi}(\chi)d\chi.
\end{align}
If $h_l$ is a complex Gaussian random variable with mean value $d_l$
and variance $\sigma^2$ and $\{h_l\}$ are mutually independent,
$\chi$ is a chi-square random variable with $2L$ degrees of freedom
and has a pdf given by
\begin{align}
f_{\chi}(\chi)=\frac{1}{\sigma^2}\left(\frac{\chi}{s^2}\right)^{\frac{L-1}{2}}e^{-\frac{\chi+s^2}{\sigma^2}}
I_{L-1}\left(\frac{2\sqrt{\chi s^2}}{\sigma^2}\right)
\end{align}
where, $s^2=\sum_{l=1}^{L}|d_l|^2$.

When the fading coefficients $\{h_l\}$ are correlated, the average
error probability $\bar{P}_e$ can be obtained by evaluating the
expected value of $P_e$ with respect to the joint distribution of
$(|h_1|, \ldots, |h_L|)$, which involves $L$-fold integration.
However, if $\{|h_l|\}$ are Nakagami-$m$ distributed, closed-form
expressions for $f_\chi(\chi)$ are provided in \cite{george1}, which
lead to a single integration.

\begin{figure}
\begin{center}
\includegraphics[width = 0.45\textwidth]{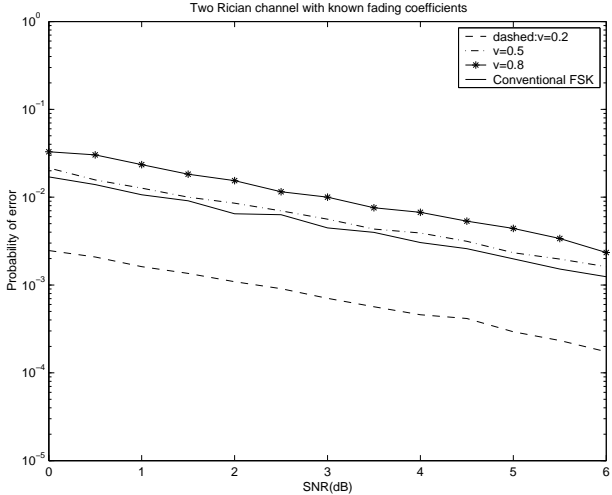}
\caption{Error probability vs. SNR for 4-OOFSK signaling over two
independent coherent Rician fading channels with equal Rician factor
$K=\frac{1}{8}$. Duty factor values are $v = 1, 0.8, 0.5$, and
$0.2$.}\label{2uncoco}
\end{center}
\end{figure}

\begin{figure}
\begin{center}
\includegraphics[width = 0.45\textwidth]{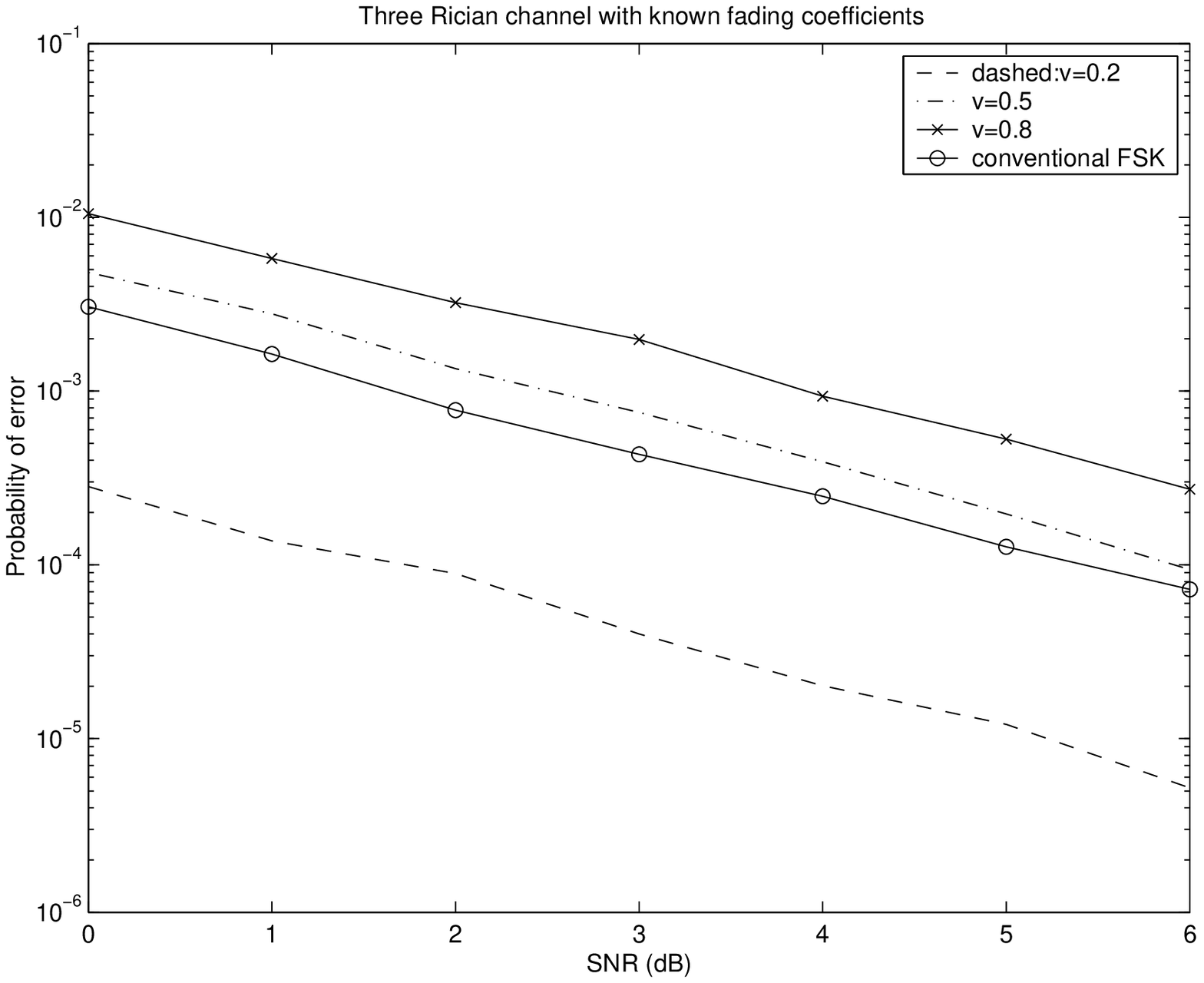}
\caption{Error probability vs. SNR for 4-OOFSK signaling over three
independent coherent Rician fading channels with equal Rician factor
$K=\frac{1}{8}$. Duty factor values are $v = 1, 0.8, 0.5$, and
$0.2$.}\label{3uncoco}
\end{center}
\end{figure}

\begin{figure}
\begin{center}
\includegraphics[width = 0.45\textwidth]{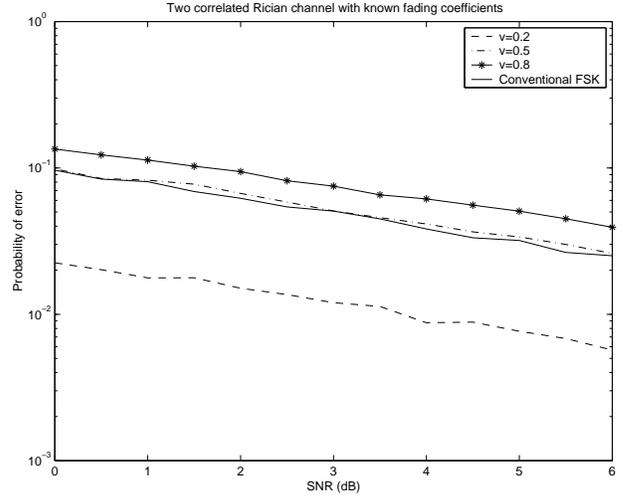}
\caption{Error probability vs. SNR for 4-OOFSK signaling over two
correlated coherent Rician fading channels with equal Rician factor
$K=\frac{1}{8}$ and correlation coefficient $\rho=\frac{1}{4}$. Duty
factor values are $v = 1, 0.8, 0.5$, and $0.2$.}\label{2coco}
\end{center}
\end{figure}

Next, we present the simulation results. We define the Rician factor
as $K=\frac{|d|^2}{\sigma^2}$ and correlation coefficient as
$\rho=\frac{\text{cov}(h_i,h_j)}{\sqrt{\text{var}(h_i)
\text{var}(h_j)}}$. Figures \ref{2uncoco}, \ref{3uncoco}, and
\ref{2coco} plot the probability of error curves as a function of
SNR for 4-OOFSK signaling over Rician fading channels with different
number of receiver antennas and different duty factors. Two
independent channels are considered in Fig. \ref{2uncoco}. Note that
conventional FSK corresponds to OOFSK with duty factor $v = 1$. In
Fig. \ref{3uncoco}, three independent channels are assumed. In both
figures, we observe an improvement in the error probability curves
if the duty factor $v$ of OOFSK signaling is less than 0.5. When $v
= 0.2$, we see approximately an order of magnitude improvement in
the error performance. This results in substantial energy gains for
fixed value of error probability, rendering OOFSK signaling a very
energy efficient transmission technique. It can also immediately be
noted that due to increased diversity, having more antennas
decreases the error rates. In Fig. \ref{2coco}, the case of two
correlated channels is investigated. Although the performance is
deteriorated due to correlation, OOFSK with sufficiently small duty
factor still considerably improves the error performance. It  should
be noted that having small duty factor means that FSK signals have
high peak power but they are transmitted less frequently to satisfy
the average power constraint. Consequently, having high peak power
signals decrease the error rates.

\section{OOFSK Over Noncoherent Fading Channels}
\label{sec:noncoherent}

\subsection{Detection Rule}

In the noncoherent channel case, we assume that the realizations of
the fading coefficients $\{h_l\}$ are unknown at both the receiver
and transmitter. The receiver is only equipped with the knowledge of
the statistics of $\{h_l\}$. We further assume that $\{h_l\}$ are
i.i.d. complex Gaussian random variables with $E\{h_l\} = d_l$ and
$var\{h_l\} = \sigma^2$. Therefore, conditioned on $s_k(t)$ being
the transmitted signal, $Y_{l,m}$ is a complex Gaussian random
variable with
\begin{align}
E\{Y_{l,m}|s_k\} &= \left\{
\begin{array}{ll}
Ad_le^{j\theta_k} & m=k
\\
0 & m\neq k
\end{array}\right.,
\nonumber \\
var\{Y_{l,m}|s_k\} &= \left\{
\begin{array}{ll}
A^2 \sigma^2 + 1 & m = k
\\
1 & m \neq k
\end{array}
\right.. \nonumber
\end{align}
Similarly as in the coherent case, we combine the energies of the
$m^\text{th}$ frequency components across the antennas, and obtain
$R_m = \sum_{l=1}^L R_{m,l}$. Conditioned on transmitted signal
$s_k(t)$, $R_m$ is a chi-square random variable with the following
pdf:
\begin{align}\label{pdf2}
f_{R_m|s_k}(R_m)\!=\!\left\{\!\!\!\!\begin{array}{ll}\frac{1}{\sigma_y^2}\left(\frac{R_m}{\xi}\right)^{\frac{L-1}{2}}
\!\!\!e^{-\frac{R_m+\xi}{\sigma_y^2}}\!\!I_{L-1}\left(\frac{2\sqrt{R_m\xi}}{\sigma_y^2}\right)
&\!\!\!m=k\\ \frac{R_m^{L-1}}{\Gamma(L)}e^{-R_m}&\!\!\! m \neq k
\end{array}\right.
\end{align}
where $\xi=A^2\sum_{l=1}^L |d_l|^2$ and $\sigma^2_y =
A^2\sigma^2+1$. The vector $\R = [R_1,\ldots,R_M]$ has the following
conditional joint pdf {\small{
\begin{align}
&f_{\R|s_k}(\R)= \nonumber \\
&\!\!\left\{\begin{array}{ll}\!\!\!\!\frac{1}{\sigma_y^2}\left(\frac{R_k}{\xi}\right)^{\frac{L-1}{2}}
\!e^{-\frac{R_k+\xi}{\sigma_y^2}}\!\!I_{L-1}\!\!\left(\frac{2\sqrt{R_k\xi}}{\sigma_y^2}\right)\!\!\prod_{n=1\
n\neq k
}^M \!\!\frac{R_n^{L-1}e^{-R_n}}{\Gamma(L)} &  \!\!\!\!k\neq 0\\
\!\!\!\frac{1}{\left[\Gamma(L)\right]^{M}}\prod_{n=1}^MR_n^{L-1}e^{-R_n}
& \!\!\!\!\!k=0\end{array}\right. \nonumber
\end{align}}}
The MAP decision rule that detects $s_k$ for $k \neq 0$ is
\begin{align}\label{compare1nc}
\left\{\begin{array}{ll}f_{\R|s_k}>f_{\R| s_m} & \forall m\neq 0, k
\\
f_{\R| s_k}>\frac{M(1-v)}{v}f_{\R| s_0}\end{array}\right.
\end{align}
Similarly as in Section \ref{sec:coherent}, it can be easily shown
that
\begin{align}\label{compare2}
g_2(R_k) =
R_k^{-\frac{L-1}{2}}e^{\frac{R_kA^2\sigma^2}{\sigma_y^2}}I_{L-1}\left(\frac{2\sqrt{R_k\xi}}{\sigma_y^2}\right),\quad
\xi>0
\end{align}
is a monotonically increasing function. With this observation, the
decision rule in (\ref{compare1nc}) simplifies to
\begin{align}\label{ncfinalcompare1}
\left\{\begin{array}{ll}R_k>R_m & \forall m \neq k\\
R_k>g_2^{-1}(T_2)\end{array}\right.
\end{align}
where
$T_2=\frac{M(1-v)\sigma_y^2\xi^{\frac{L-1}{2}}e^{\frac{\xi}{\sigma_y^2}}}{v\Gamma(L)}.$
Note that $s_0$ is the detected signal if $R_k < g_2^{-1}(T_2)$ for
all $k$.

\subsection{Probability of Error}
\begin{figure}
\begin{center}
\includegraphics[width = 0.45\textwidth]{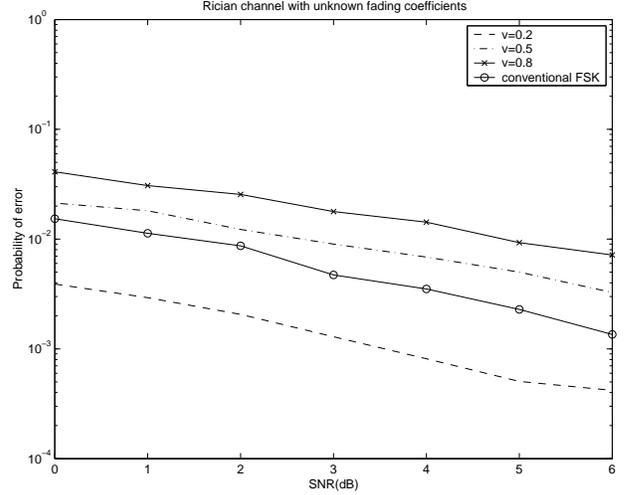}
\caption{Error probability vs. SNR for 4-OOFSK signaling over two
independent noncoherent Rician fading channels with equal Rician
factor $K=\frac{1}{8}$.}\label{2non}
\end{center}
\end{figure}

\begin{figure}
\begin{center}
\includegraphics[width = 0.45\textwidth]{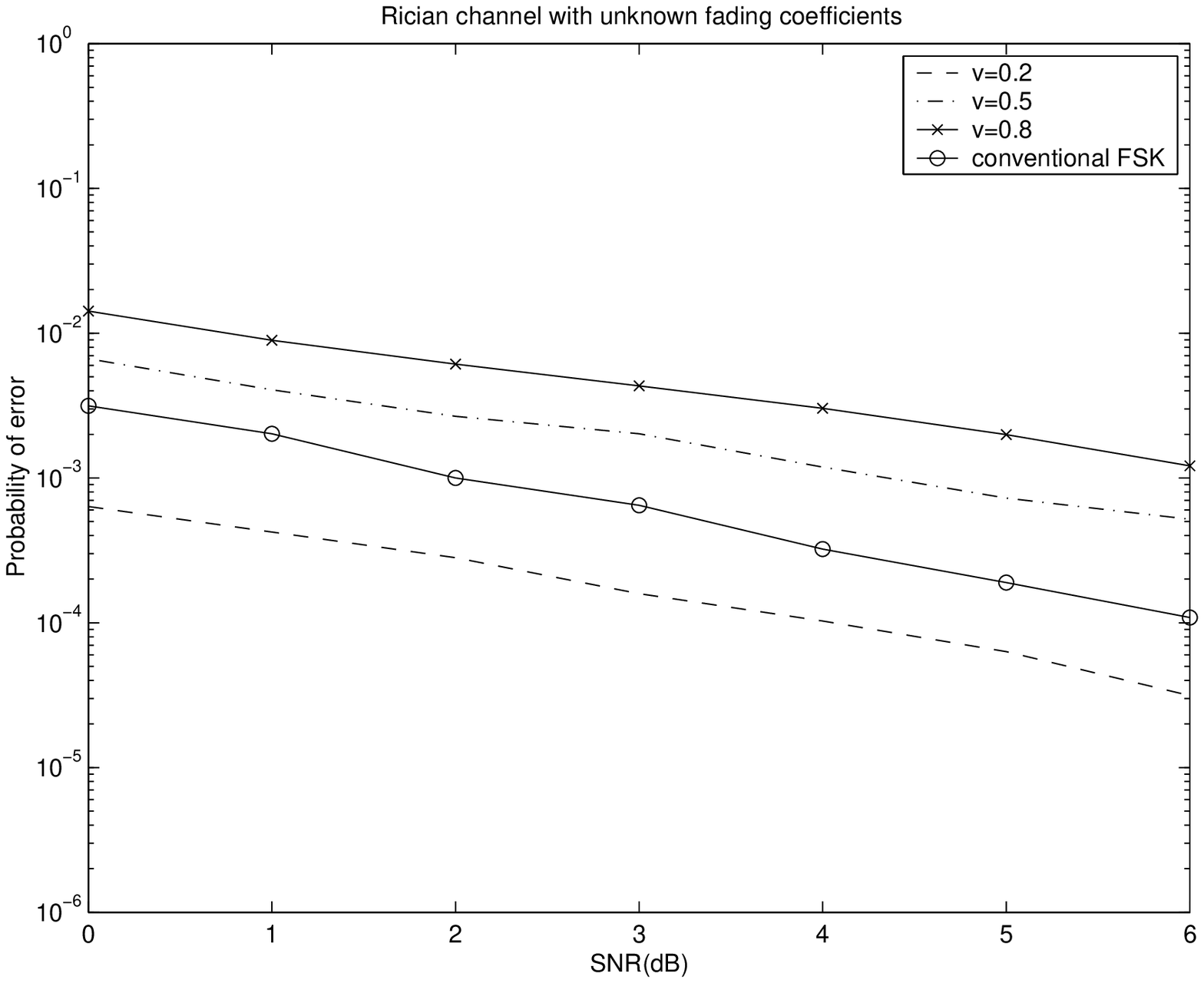}
\caption{Error probability vs. SNR for 4-OOFSK signaling over three
indepedent noncoherent Rician fading channels with equal Rician
factor $K=\frac{1}{8}$.}\label{3non}
\end{center}
\end{figure}

\begin{figure}
\begin{center}
\includegraphics[width = 0.45\textwidth]{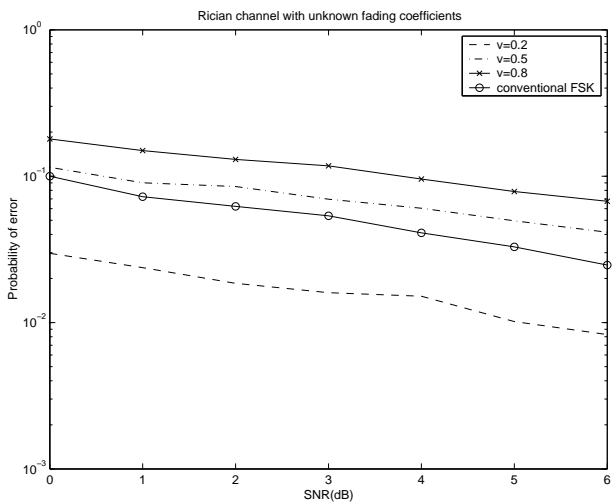}
\caption{Error probability vs. SNR for 4-OOFSK signaling over two
correlated noncoherent Rician fading channels with equal Rician
factor $K=\frac{1}{8}$.} \label{3noncor}
\end{center}
\end{figure}

We first assume that $s_1(t)$ is transmitted. Let $\tau_2 =
g_2^{-1}(T_2)$. Then, the probability of correct detection is
\begin{align}
P_{c,1}\!\!=\!\!P(R_2\!>\!R_1,R_3\!>\!R_1,\ldots,R_M\!>\!R_1,R_1\!>\!\tau_2|s_1)
\end{align}
Following an approach similar to that in Section \ref{sec:coherent},
we have
\begin{align}
P_{c,1}&\!\!=\!\!\sum_{n=0}^{M-1}(-1)^n\left(\!\!\!\begin{array}{cc}M-1\\n\end{array}\!\!\!\right)
         \!\!\!\sum_{i=0}^{n(L-1)}\!\!c_{in}\int_{\tau_2}^\infty\!\!x^ie^{-nx}f_{R_1|s_1}(x)dx\nonumber\\
&=\sum_{n=0}^{M-1}(-1)^n\left(\begin{array}{cc}M-1\\n\end{array}\right)
\sum_{i=0}^{n(L-1)}c_{in}\int_{\tau_2}^\infty x^ie^{-nx}\nonumber\\
&\quad
\times\frac{1}{\sigma_y^2}\left(\frac{x}{\xi}\right)^{\frac{L-1}{2}}e^{-\frac{x+\xi}{\sigma_y^2}}I_{L-1}\left(\frac{2\sqrt{x\xi}}{\sigma_y^2}\right)dx
\nonumber\\
&=\sum_{n=0}^{M-1}(-1)^n\left(\begin{array}{cc}M-1\\n\end{array}\right)\sum_{i=0}^{n(L-1)}c_{in}
\frac{\xi^{-\frac{L-1}{2}}e^{-\frac{\xi}{\sigma_y^2}}}{\sigma_y^2}\nonumber\\
&\times
\left[\frac{\xi^{\frac{L-1}{2}}(i+L)!}{2(1+n\sigma_y^2)^{\frac{i+L}{2}}
\sigma_y^{L-2-i}L!}F\left(-i,L,\frac{\xi}{\sigma_y^2(1+n\sigma_y^2)}\right)\right.\nonumber\\
&\left.\times
e^{\frac{\xi}{\sigma_y^2(1+n\sigma_y^2)}}-\int_0^{\tau_2}
x^{\frac{2i+L-1}{2}}e^{-\frac{1+n\sigma_y^2}{\sigma_y^2}x}I_{L-1}\left(\frac{2\sqrt{x\xi}}{\sigma_y^2}\right)dx\right]
\nonumber
\end{align}
If $s_0(t)$ is the transmitted signal, the probability of correct
detection is
\begin{align}
P_{c,0}&=P\left(R_1<\tau_2,\ldots,R_M<\tau_2 |s_0\right)\nonumber\\
&= \left( 1 - e^{-\tau} \sum_{l=0}^{L-1}
\frac{\tau^l}{l!}\right)^M.
\end{align}
Finally, the average probability of error is
\begin{align}
P_e=1-(vP_{c,1}+(1-v)P_{c,0}).
\end{align}
Figures \ref{2non}, \ref{3non}, and \ref{3noncor} provide the
simulation results of error probability when 4-OOFSK signals are
transmitted over noncoherent Rician fading channels. In Figs.
\ref{2non} and \ref{3non}, the channels are assumed to be
independent. In these figures, it is seen that OOFSK signaling with
$v =0.8$ and $v = 0.5$ have worse error performance when compared to
that of conventional FSK (OOFSK with $v = 1$). As evidenced in the
graph of $v = 0.2$, if the duty factor is sufficiently decreased,
and hence consequently the peak power is increased, we start seeing
improvements. Since fading is not known in the noncoherent case, the
advantage of using OOFSK signaling is twofold. Having low duty cycle
allows the FSK signals to have high peak power which is especially
beneficial when channel characteristics are unknown. In addition,
when the zero signal $s_0(t)$ is sent, the received signal is
composed of additive noise and is free of fading coefficients.
Finally, Fig. \ref{3noncor} plots the error probabilities when
4-OOFSK signals are sent over two correlated noncoherent channels.
These curves are obtained when the detection rule
(\ref{ncfinalcompare1}) derived for independent channels are
employed at the receiver. We note that the performance degrades due
to correlation in noncoherent channels as well. Similar conclusions
about OOFSK modulation are drawn.

\end{document}